\pdfoutput=1

\documentclass[11pt]{article}

\usepackage[preprint]{acl}

\usepackage{times}
\usepackage{latexsym}

\usepackage[T1]{fontenc}

\usepackage[utf8]{inputenc}

\usepackage{microtype}

\usepackage{inconsolata}

\usepackage{graphicx}

\title{Evaluating Named Entity Recognition Using Few-Shot Prompting with Large Language Models}

\author{
   \textbf{Hédi Zeghidi} \and
   \textbf{Ludovic Moncla}
\\
\\
  INSA Lyon, CNRS, Universite Claude Bernard Lyon 1,\\LIRIS, UMR5205, 69621 Villeurbanne, France 
\\
  \small{
    \textbf{Correspondence:} \href{mailto:ludovic.moncla@insa-lyon.fr}{ludovic.moncla@insa-lyon.fr}
  }
}

\begin{document}

\maketitle

\begin{abstract}
This paper evaluates Few-Shot Prompting with Large Language Models for Named Entity Recognition (NER). Traditional NER systems rely on extensive labeled datasets, which are costly and time-consuming to obtain. Few-Shot Prompting or in-context learning enables models to recognize entities with minimal examples. We assess state-of-the-art models like GPT-4 in NER tasks, comparing their few-shot performance to fully supervised benchmarks. Results show that while there is a performance gap, large models excel in adapting to new entity types and domains with very limited data. We also explore the effects of prompt engineering, guided output format and context length on performance. This study underscores Few-Shot Learning's potential to reduce the need for large labeled datasets, enhancing NER scalability and accessibility.
\end{abstract}

\section{Introduction}

Named Entity Recognition (NER) is a fundamental task in Natural Language Processing (NLP), involving the identification and classification of entities such as names, locations, and dates within text. Traditional NER systems typically require large, annotated datasets for effective training, which can be expensive and time-consuming to curate. This limitation hinders the rapid deployment and scalability of NER applications across diverse domains and languages.
Few-Shot Learning (FSL) offers a promising alternative by enabling models to perform NER with minimal labeled examples. Leveraging the capabilities of Large Language Models (LLMs), FSL can significantly reduce the dependency on extensive annotated data. This paper investigates the performance of LLMs in NER tasks using FSL, comparing their efficacy to conventional fully supervised methods.
Through this evaluation, we aim to highlight the potential of FSL to transform the landscape of NER, making it more accessible and scalable.

Despite their impressive performance in many NLP tasks, NER remains challenging for LLMs \cite{lu2024large, gonzalez2024leveraging}. Mainly because NER is inherently a sequence labeling task, whereas LLMs are primarily designed for text generation.
Existing methods utilizing LLMs typically extract only a list of entities from text along with their respective classes. 
We aim to improve the process by identifying each entity's position in the text and outputting it in JSON format, including both token and character positions.

This study aims to answer the following question: 
How effective are FSL techniques in LLMs for performing NER tasks?
Our experiments are perform using the GeoEDdA dataset\footnote{\url{https://huggingface.co/datasets/GEODE/GeoEDdA}} \cite{moncla2024geoedda} and several LLMs are evaluated and compared to BERT-like model \cite{devlin2018bert} baselines.

\section{Related works}
\label{related_works}

Some works study the zero-shot capabilies of GPT for NER. For instance, \citet{xie2023empirical} proposed the NER task decomposition into a set of simpler subproblems by labels and perform a decomposed-question-answering task in addition to a syntactic augmentation.
However, most of works are investigating FSL. 
\citet{li2023far} investigate the performances of LLMs and smaller pre-trained language models (e.g., BERT, T5) for domain-specific NER. 
\citet{ashok2023promptner} propose to prompts an LLM to produce a list of potential entities along with corresponding explanations justifying their compatibility with the provided entity type definitions. 
\citet{wang2023gpt} propose adding special tokens for marking the entities to extract. The evaluation shows that this method achieve comparable results as supervised methods.

For newly models with expanded context windows, \cite{agarwal2024many} propose an experimental study comparing FSL versus many-shot learning (hundreds or thousands of examples). Results indicate significant performance improvements when increasing the number of provided examples from very few to many. Results also indicate that the order of examples in the prompt affects many-shot performance and adding more examples than optimal can sometimes degrade performance for certain tasks. Additionally, providing only questions without answers in the examples works well in some configurations. These observations are model-dependent, based on tests conducted with Gemini 1.5, GPT-4, and Claude 3.5.

\section{Methodology}
\label{methodology}

The objective of this preliminary work is to evaluate the capabilities of LLMs on the NER task at the token and span (or entity) levels. For these two tasks, the output must be a JSON format with information for each detected token or span. A span refer to an entity and can be composed of several tokens.

Our methodology relies on a prompt containing descriptions of the task, tagset and one example (with both input and output). 
A preliminary experiment reveals that LLMs struggle to accurately retrieve or calculate token or span positions from raw input text. 
Although the output format was correct, the generated position values were inaccurate, resembling hallucinations or random numbers rather than referring to actual token positions, despite appearing as numerical values.
To address this issue, we propose a solution that includes tokenization information—specifically, token position details for span detection—within the input data.

\section{Experiments and Results}
\label{experiments}

The experiments\footnote{Source code publicly available at \url{https://github.com/GEODE-project/ner-llm}} are performed using the GeoEDdA dataset \cite{moncla2024geoedda} which contains semantic annotations (at the token and span levels) for named entities (i.e., Spatial, Person, and Misc), nominal entities, spatial relations, and geographic coordinates. Nested named entities \cite{gaio2017extended} also present in this dataset were not considered in this experiment.
One example\footnote{\url{https://huggingface.co/datasets/GEODE/GeoEDdA/viewer/default/train?row=53}} from the training set containing at least one entity of each class is used for FSL (included in the prompt) and the entire test set (200 articles from the Diderot Encyclopedie) is used for evaluation.

GPT models from the OpenAI\footnote{\url{https://openai.com}} API (gpt-3.5-turbo-0125, gpt-4-0613 and gpt-4o-2024-05-13) were evaluated through the LangChain Python framework. 
Token-level scores are shown in Table~\ref{tab:token_scores}. Micro average precision, recall and F1-score are used as evaluation metrics in a strict matching set (exact boundary matching). 
While the performance is below that of a traditionally fine-tuned BERT model\footnote{\url{https://huggingface.co/GEODE/bert-base-french-cased-edda-ner}} (fine-tuned using the fully supervised training set containing 1,800 encyclopaedia entries), experiments have shown significant variation across the different classes. 
At the span level, some unusual issues (compared to traditional NER) arise as the output format is more complex than at the token level. 
The experiment confirmed the performance improvement between GPT versions 4 and 3.5. Using GPT-3.5, 28\% of the predicted spans only are correct (both boundaries and labels) compared to 49\% with GPT-4o, while 13\% are partially correct (partial boundaries and correct labels) with GPT-3.5 and 9\% with GPT-4o
Also, using GPT-3.5, some answers do not refer to the input document but rather to the example from the prompt. In very few cases, some answers do not strictly follow the predefined JSON output format and some token attributes may be missing (such as start, end, text, or label). 


\begin{table}
  \centering
  \begin{tabular}{lcccc}
    \hline
    \textbf{Model} & \textbf{Precision} & \textbf{Recall}    & \textbf{F1} \\
    \hline
GPT-3.5			& 0.81     		& 0.36	    & 0.50   \\  
GPT-4			& 0.75     		& 0.62     	& 0.67   \\  
GPT-4o	 		& 0.68     		& 0.72     	& 0.70   \\ \hline
Fine-tuned BERT & 0.93		    & 0.94		& 0.93   \\ \hline 
  \end{tabular}
  \caption{Token classification scores}
  \label{tab:token_scores}
\end{table}

We also experimented smaller and local LLMs using LM Studio\footnote{\url{https://lmstudio.ai}} such as Phi3 (Microsoft), Gemma (Google), Mistral (MistralAI), Qwen (Qwen Team, affiliated
with Alibaba Group), and Llama (Meta). 
The models were executed on an Nvidia RTX 3500 ADA GPU and obtained very varying results. Some LLMs provide the correct JSON output syntax but don’t really understand the defined set of labels. Others completely don’t understand the task and do anything or simply repeat the input sentence.

\section*{Acknowledgments}

The authors are grateful to the ASLAN project (ANR-10-LABX-0081) of the Université de Lyon, for its financial support within the French program "Investments for the Future" operated by the National Research Agency (ANR).

\bibliography{acl_latex}

\begin{thebibliography}{10}
\providecommand{\natexlab}[1]{#1}

\bibitem[{Agarwal et~al.(2024)Agarwal, Singh, Zhang, Bohnet, Chan, Anand, Abbas, Nova, Co-Reyes, Chu et~al.}]{agarwal2024many}
Rishabh Agarwal, Avi Singh, Lei~M Zhang, Bernd Bohnet, Stephanie Chan, Ankesh Anand, Zaheer Abbas, Azade Nova, John~D Co-Reyes, Eric Chu, et~al. 2024.
\newblock \href {https://arxiv.org/abs/2404.11018} {Many-shot in-context learning}.
\newblock \emph{arXiv preprint arXiv:2404.11018}.

\bibitem[{Ashok and Lipton(2023)}]{ashok2023promptner}
Dhananjay Ashok and Zachary~C Lipton. 2023.
\newblock \href {https://arxiv.org/abs/2305.15444} {Promptner: Prompting for named entity recognition}.
\newblock \emph{arXiv preprint arXiv:2305.15444}.

\bibitem[{Devlin et~al.(2018)Devlin, Chang, Lee, and Toutanova}]{devlin2018bert}
Jacob Devlin, Ming-Wei Chang, Kenton Lee, and Kristina Toutanova. 2018.
\newblock \href {https://arxiv.org/abs/1810.04805} {Bert: Pre-training of deep bidirectional transformers for language understanding}.
\newblock \emph{arXiv preprint arXiv:1810.04805}.

\bibitem[{Gaio and Moncla(2017)}]{gaio2017extended}
Mauro Gaio and Ludovic Moncla. 2017.
\newblock Extended named entity recognition using finite-state transducers: An application to place names.
\newblock In \emph{The ninth international conference on advanced geographic information systems, applications, and services}, Nice, France.

\bibitem[{Gonz{\'a}lez-Gallardo et~al.(2024)Gonz{\'a}lez-Gallardo, Hanh, Hamdi, and Doucet}]{gonzalez2024leveraging}
Carlos-Emiliano Gonz{\'a}lez-Gallardo, Tran Thi~Hong Hanh, Ahmed Hamdi, and Antoine Doucet. 2024.
\newblock Leveraging open large language models for historical named entity recognition.
\newblock In \emph{The 28th International Conference on Theory and Practice of Digital Libraries}, Ljubljana, Slovenia.

\bibitem[{Li and Zhang(2023)}]{li2023far}
Mingchen Li and Rui Zhang. 2023.
\newblock \href {https://arxiv.org/abs/2307.00186} {How far is language model from 100\% few-shot named entity recognition in medical domain}.
\newblock \emph{arXiv preprint arXiv:2307.00186}.

\bibitem[{Lu et~al.(2024)Lu, Li, Wen, Wang, Wang, and Liu}]{lu2024large}
Qiuhao Lu, Rui Li, Andrew Wen, Jinlian Wang, Liwei Wang, and Hongfang Liu. 2024.
\newblock \href {https://arxiv.org/abs/2407.00731} {Large language models struggle in token-level clinical named entity recognition}.
\newblock \emph{arXiv preprint arXiv:2407.00731}.

\bibitem[{Moncla et~al.(2024)Moncla, Vigier, and Mcdonough}]{moncla2024geoedda}
Ludovic Moncla, Denis Vigier, and Katherine Mcdonough. 2024.
\newblock {GeoEDdA}: {A Gold Standard Dataset for Geo-semantic Annotation of Diderot \& d'Alembert's Encyclop{\'e}die}.
\newblock In \emph{Second International Workshop on Geographic Information Extraction from Texts (GeoExT)}, Glasgow, Scotland.

\bibitem[{Wang et~al.(2023)Wang, Sun, Li, Ouyang, Wu, Zhang, Li, and Wang}]{wang2023gpt}
Shuhe Wang, Xiaofei Sun, Xiaoya Li, Rongbin Ouyang, Fei Wu, Tianwei Zhang, Jiwei Li, and Guoyin Wang. 2023.
\newblock \href {https://arxiv.org/abs/2304.10428} {Gpt-ner: Named entity recognition via large language models}.
\newblock \emph{arXiv preprint arXiv:2304.10428}.

\bibitem[{Xie et~al.(2023)Xie, Li, Zhang, Zhang, Liu, and Wang}]{xie2023empirical}
Tingyu Xie, Qi~Li, Jian Zhang, Yan Zhang, Zuozhu Liu, and Hongwei Wang. 2023.
\newblock \href {https://arxiv.org/abs/2310.10035} {Empirical study of zero-shot ner with chatgpt}.
\newblock In \emph{Proceedings of the 2023 Conference on Empirical Methods in Natural Language Processing}, pages 7935--7956, Singapore.

\end{thebibliography}

\end{document}